\documentclass[prl,twocolumn,superscriptaddress,fixfloat]{revtex4}

\usepackage{amsmath}
\usepackage{amssymb}
\usepackage{graphicx}

\begin{document}

\title{Anomalous High-Energy Spin Excitations in La$_2$CuO$_4$}

\author{N. S. Headings}
\affiliation{H.H. Wills Physics Laboratory, University of Bristol, Tyndall Ave., Bristol, BS8 1TL, UK}
\author{S. M. Hayden}
\email{s.hayden@bris.ac.uk}
\affiliation{H.H. Wills Physics Laboratory, University of Bristol, Tyndall Ave., Bristol, BS8 1TL, UK}
\author{R. Coldea}
\affiliation{H.H. Wills Physics Laboratory, University of Bristol, Tyndall Ave., Bristol, BS8 1TL, UK}
\affiliation{Clarendon Laboratory, Department of Physics, University
of Oxford, Oxford OX1 3PU, United Kingdom}
\author{T. G. Perring}
\affiliation{ISIS Facility, Rutherford Appleton Laboratory, Chilton, Didcot, Oxfordshire OX11 0QX, UK}

\begin{abstract}
Inelastic neutron scattering is used to investigate the collective magnetic excitations of the high-temperature superconductor parent antiferromagnet La$_2$CuO$_4$. We find that while the lower energy excitations are well described by spin-wave theory, including one- and two-magnon scattering processes, the high-energy spin waves are strongly damped near the (1/2,0) position in reciprocal space and merge into a momentum dependent continuum. This anomalous damping indicates the decay of spin waves into other excitations, possibly unbound spinon pairs.
\end{abstract}

\maketitle

High-temperature superconductivity occurs when insulating layered antiferromagnets such as La$_2$CuO$_4$ and YBa$_2$Cu$_3$O$_{6.15}$ are doped with electrons or holes \cite{Lee2006a}.  La$_2$CuO$_4$ and the other parent compounds of the high-temperature superconductors are Mott-insulating layered square lattice antiferromagnets with large exchange couplings. Doping these parent antiferromagnets results in a collapse of the antiferromagnetic (AF) order, a metal insulator transition and the formation of a superconducting phase with a high transition temperature.  Many believe that the global phase diagram of the cuprates is explained in a strong coupling theory in which the spin correlations important for superconductivity are already present in the undoped parent compounds \cite{Anderson2004a,Lee2006a,Anderson1987a}.
For example, Anderson \cite{Anderson1987a} proposed that the insulating parent compounds are described by a ``resonating-valence-bond'' (RVB) model.  In this model, the unpaired electrons on the Cu atoms pair with neighbours and are described by a superposition of singlet pairs.  Hole doping introduces vacancies which can hop in a background of singlets. The pre-existing magnetic singlets may become the charged superconducting pairs in the superconductor \cite{Anderson2004a,Lee2006a}.

In this paper, we use inelastic neutron scattering (INS) to measure the collective magnetic excitations of the insulating parent compound La$_2$CuO$_4$ over a wide range of energy and momentum.  Our experiments were performed using the MAPS spectrometer at the ISIS spallation neutron source of the Rutherford-Appleton Laboratory.  The use of a spectrometer with a large angular detector coverage and crystals with good mosaic, has enabled considerable improvements on previous studies \cite{Coldea2001a}.  Thus we are able to observe new contributions to the magnetic response which carry significant spectral weight and highlight the strong fluctuations present in La$_2$CuO$_4$ at high energies.
\begin{figure}
\includegraphics[width=0.75\linewidth,clip]{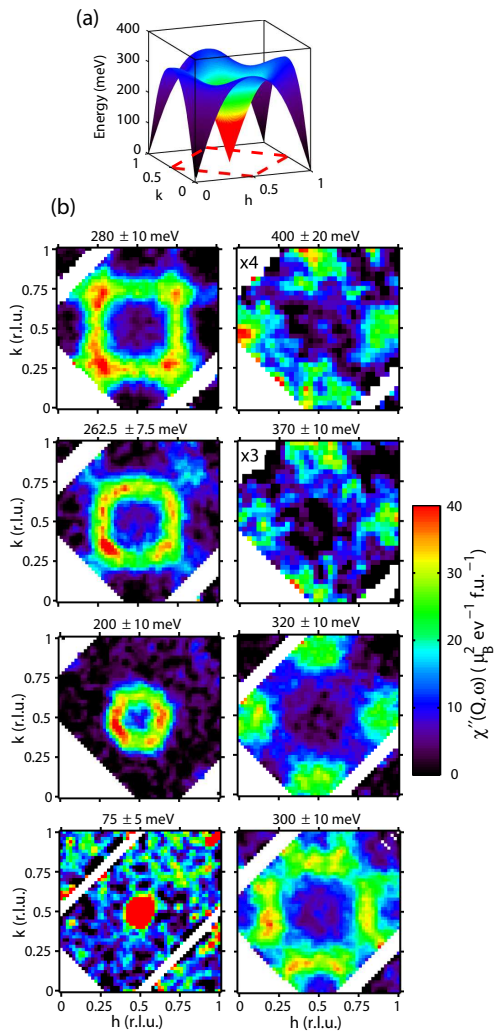}
\caption{(Color online) (a) Dispersion as predicted by SWT (color indicates the scattering intensity). (b) The measured $\chi^{\prime\prime}(\mathbf{q},\omega)$ in units of $\mu_B^2$ eV$^{-1}$ f.u.$^{-1}$ at $T=10$~K. At the highest energies ($E$$\geq$320~meV), the response is strongest near $\mathbf{q}=(1/2,0)$ and symmetry related positions.}
\label{Fig:q_slices}
\end{figure}
\begin{figure}
\includegraphics[width=0.7\linewidth,clip]{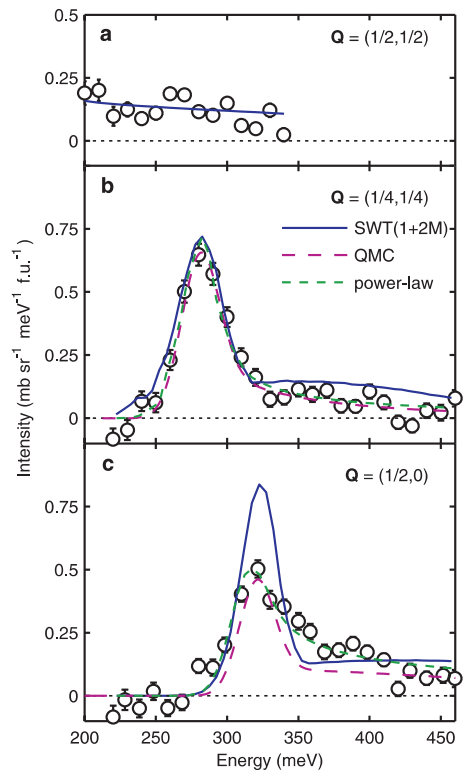}
\caption{(Color online) Magnetic excitation spectrum at various wavevectors in La$_{2}$CuO$_{4}$. (a) Ordering wavevector showing the two-magnon continuum. (b) (1/4,1/4) position on the magnetic zone boundary. (c) (1/2,0) position where the spin wave pole is strongly damped (zone boundary anomaly). Lines show resolution-convolved fits to: a spin wave theory one- plus two-magnon cross-section [SWT(1+2M)]; quantum Monte Carlo simulation (QMC); and a generic power law continuum lineshape (Eq.~\ref{Eq:Spinon}). Data have been averaged over equivalent points for $h,k \leq 1$ and a background at (1,0) has been subtracted.}
\label{Fig:e_cuts}
\end{figure}
\begin{figure}
\includegraphics[width=0.8\linewidth,clip]{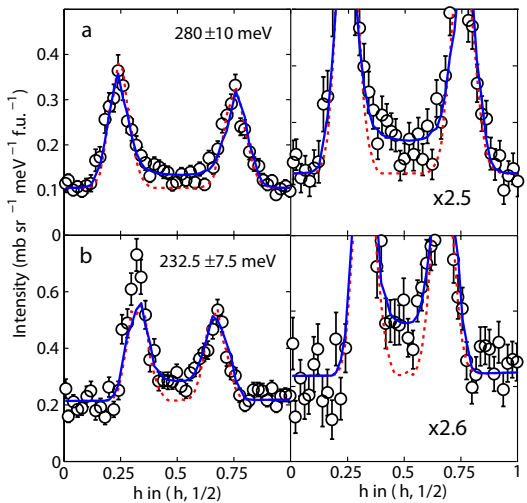}
\caption{(Color online) Spin-wave and continuum (two-magnon) scattering in La$_2$CuO$_4$.
(a)-(b) Cuts along $\mathbf{Q}=(h, 0.5)$ for energy transfers $\hbar\omega$=232.5 and 280~meV and integration ranges shown. The dotted line is a fit of  SWT (one-magnon only) convolved the instrumental resolution with a quadratic background.  The solid line is a fit of one-magnon plus two-magnon model.  Right panels are magnified.}
\label{Fig:twomag_q_cuts}
\end{figure}

\begin{figure*}
\includegraphics[width=0.85\linewidth,clip]{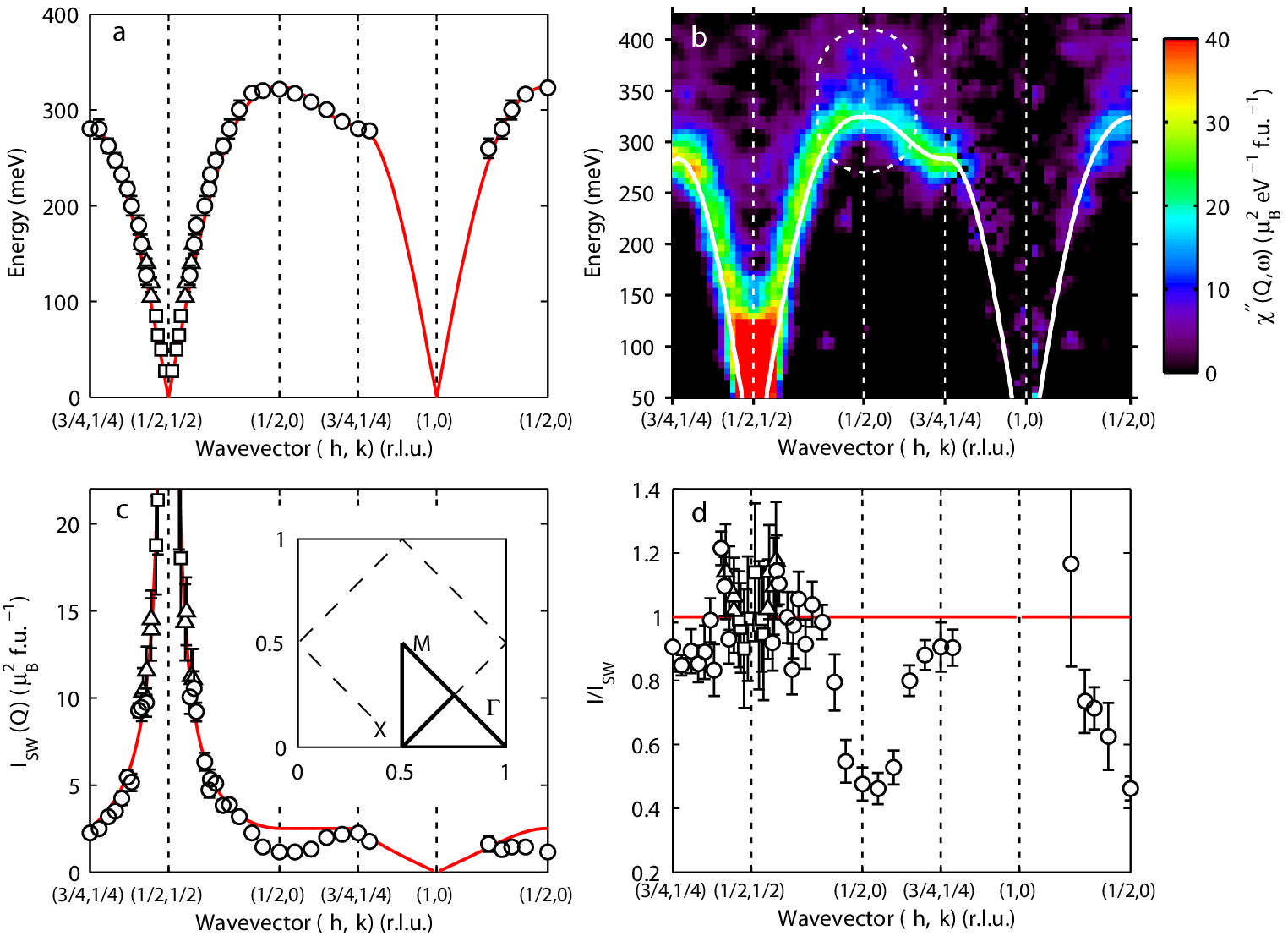}
\caption{(Color online) Wavevector dependence of the magnetic excitations in La$_2$CuO$_4$ showing anomalous response near (1/2,0).
(a) One-magnon dispersion ($T=10$~K) along lines in Fig.~\ref{Fig:dispersion}(c) (inset). Symbols indicate $E_i$: 160 meV ($\Box$), 240 meV ($\triangle$) and 450 meV ($\circ$).  The solid line is a SWT fit based on Eq.~\ref{Eq:Heisenberg}.  (b) Measured $\chi^{\prime\prime}(\mathbf{q},\omega)$ throughout the Brillouin zone. Dashed circled highlights the anomalous scattering near (1/2,0). An energy-dependent background determined near (1,0) has been subtracted.  (c) Experimentally-extracted one-magnon intensity. Line is a fit to SWT with renormalization factor $Z_d$=0.4$\pm$$0.04$. Dashed line in inset shows the AF Brillouin zone. (d) One-magnon intensity divided by SWT prediction highlighting anomalous loss in intensity at (1/2,0).}
\label{Fig:dispersion}
\end{figure*}
The collective magnetic excitations in La$_2$CuO$_4$ are due to correlations between the $S=1/2$ unpaired spins associated with the Cu atoms. INS probes the spin-spin correlations as a function of wavevector $\mathbf{q}$ and energy $\hbar\omega$. The neutron cross-section may be directly related to the imaginary part of the magnetic susceptibility $\chi^{\prime \prime}(\mathbf{q},\omega)$, which is a measure of the strength of the excitations for a given $(\mathbf{q},\omega)$.  We specify wavevectors in terms of their positions in reciprocal space $\mathbf{Q}=h \mathbf{a}^{\star}+k \mathbf{b}^{\star}+ l \mathbf{c}^{\star}$ of the tetragonal (or square) lattice, where $a \approx 3.8$ and $c \approx13.1$~\AA.  Our sample consisted of five co-aligned single crystals of La$_2$CuO$_4$, with total mass of 47.5~g.  The crystals were annealed in Argon at 1073~K for 48h.  Susceptibility measurements showed a N\'{e}el temperature of $320 \pm 5$~K. INS data were collected at 10 K with incident energies of 160, 240, 450 and 600 meV and corrected using the Cu$^{2+}$ $d_{x{\scriptstyle ^2}-y{\scriptstyle ^2}}$ form factor.

Fig.~\ref{Fig:q_slices} shows typical data in the form of constant energy slices plotted as a function of in-plane wavevector $(h,k)$.  The high-intensity regions in Fig.~\ref{Fig:q_slices} are produced when the spin-wave (SW) dispersion surface $\omega(\mathbf{q})$ (Fig.~\ref{Fig:q_slices}(a)) intersects with the measuring plane.  Fig.~\ref{Fig:e_cuts} and Fig.~\ref{Fig:twomag_q_cuts} show energy- and wavevector-dependent cuts through the excitations which allow the spin-wave pole to be located.  From these and other cuts, we are able to determine the SW dispersion throughout the Brillouin zone.  This is shown in Fig.~\ref{Fig:dispersion}.  The results for the SW dispersion are similar to those obtained in previous INS and RIXS studies \cite{Coldea2001a,Braicovich2010a}.

An approach which has been widely used to describe the magnetic excitations in La$_2$CuO$_4$ is to assume that: (i) the magnetic interactions are described by a Heisenberg model on a square lattice; (ii) the ground state has N\'{e}el order; and (iii) the excitations can be computed using classical (large-$S$) linear spin-wave theory (SWT). Following previous work \cite{Coldea2001a} we used a Heisenberg Hamiltonian with higher order coupling to model the magnetic interactions:
\begin{eqnarray}
\label{Eq:Heisenberg}
{\cal H} & = & J \sum_{\langle i,j \rangle} {\bf S}_i \cdot {\bf S}_j +
 J^{\prime} \sum_{\langle i,i^{\prime} \rangle} {\bf S}_i \cdot {\bf S}_{i^{\prime}} +
 J^{\prime\prime} \sum_{\langle i,i^{\prime\prime} \rangle} {\bf S}_i \cdot {\bf S}_{i^{\prime\prime}} \nonumber \\
& &
+ J_c \sum_{\langle i,j,k,l \rangle}
\left\{ ({\bf S}_i \cdot {\bf S}_j)({\bf S}_k \cdot {\bf S}_l) +
({\bf S}_i \cdot {\bf S}_l)({\bf S}_k \cdot {\bf S}_j) \right. \nonumber \\
& &
- \left. ({\bf S}_i \cdot {\bf S}_k)({\bf S}_j \cdot {\bf S}_l)\right\},
\end{eqnarray}
where $J$, $J^{\prime}$ and $J^{\prime\prime}$ are the first-, second- and third-nearest-neighbour exchange constants and $J_c$ is the ring exchange interaction coupling four spins. The exchange constants in Eq.~\ref{Eq:Heisenberg} can be estimated in a Hubbard model in terms of the hopping energy $t$ and double occupancy energy $U$ \cite{MacDonald1900a,Coldea2001a,Delannoy2009a}. By fitting $t$ and $U$, we obtained $J$=143$\pm$2~meV, $J^{\prime}$=$J^{\prime\prime}$=2.9$\pm$0.2~meV and $J_c$=58$\pm$4~meV. The resulting dispersion surface is shown in Fig.~\ref{Fig:q_slices}(a) and dispersion curves in Fig.~\ref{Fig:dispersion}(a). A slightly improved fit to the intermediate-energy dispersion points can be obtained by including higher order hopping or exchange terms as in Refs.~\cite{Delannoy2009a,Guarise2010a}.

The SW dispersion has been measured in detail in a number of $S=1/2$ square lattice Heisenberg antiferromagnets \cite{Kim1999a,Ronnow2001a,Coldea2001a,Christensen2007a,Tsyrulin2009a,Guarise2010a}.  The dispersion along the zone boundary between (3/4,1/4) and (1/2,0) is subject to two competing effects. For a nearest neighbor only coupling, SWT predicts there is no dispersion.  Quantum fluctuations, not taken into account by SWT, cause the one-magnon energies to be renormalized raising (3/4,1/4) with respect to (1/2,0) \cite{Sandvik2001a}. This is observed experimentally when $t/U$ (and $J$) is small \cite{Kim1999a,Ronnow2001a,Christensen2007a,Tsyrulin2009a}. For larger $t/U$ (e.g. La$_2$CuO$_4$) \cite{Coldea2001a,Guarise2010a}, the opposite happens as the higher order exchange coupling ($J_c$) in Eq.~\ref{Eq:Heisenberg} raises (1/2,0) with respect to (3/4,1/4).

SWT also makes a prediction about the $\mathbf{q}$-dependence of the intensity of the SW pole.  The solid line in Fig.~\ref{Fig:dispersion}(c)-(d) shows this SWT prediction compared to the integrated intensity obtained from our data. Overall there is good agreement with the SWT prediction expect near $\mathbf{q}=(1/2,0)$ [or $(\pi,0)$], where the SW peak intensity is strongly suppressed.  Inspection of the energy-dependent scans Fig.~\ref{Fig:e_cuts}(b)-(c) shows that in the region of $\mathbf{q}$-space near $(1/2,0)$, where the SW pole is suppressed, there is a significant high-energy tail to the magnetic response. This is also visible in the $\mathbf{q}$-dependent plots Fig.~\ref{Fig:q_slices} for $E$$\geq$320~meV, which show the persistence of the scattering well localized near $(1/2,0)$ and symmetry related points to the highest energies investigated. The high energy tail and intensity modulation can be seen most dramatically in the image of the magnetic excitations along the high-symmetry directions shown in Fig.~\ref{Fig:dispersion}(b).  Here, the tail at (1/2,0) and the intensity modulation between (1/2,0) and (3/4,1/4) can be seen. We also see evidence for continuum scattering at lower energies inside the spin-wave cone as illustrated by the scans in Fig.~\ref{Fig:twomag_q_cuts}.  In an antiferromagnet with collinear N\'{e}el order, a two-magnon continuum \cite{Heilmann1981a,Lorenzana2005a,Christensen2007a} is expected due to the coherent creation of two independently-propagating spin waves with opposite spin ($S_z=\pm 1$). This has been observed in other square-lattice antiferromagets such as Rb$_{2}$MnF$_{4}$ \cite{Huberman2005a} and copper deuteroformate tetradeurate (CFTD) \cite{Christensen2007a}.  We have computed the two-magnon cross section based on the spin-wave dispersion determined by Eq.~\ref{Eq:Heisenberg} using the standard expression \cite{Lorenzana2005a,Heilmann1981a} (see solid lines in Figs.~\ref{Fig:e_cuts} and \ref{Fig:twomag_q_cuts}).  While the two-magnon simulation gives a good overall description of the extra low-energy continuum inside the spin-wave cone, it fails to account for the high-energy lineshape. Inspection of Fig.~\ref{Fig:e_cuts}(a)-(c) shows that the response at $\mathbf{q}=(1/4,1/4)$ is overestimated and that at $\mathbf{q}=(1/2,0)$ underestimated.

We also compared our data with quantum Monte Carlo (QMC) simulations \cite{Sandvik2001a,note_QMC} which are shown as a dashed line in Fig.~\ref{Fig:e_cuts}. The QMC simulation provides a good description of the $\mathbf{q}$-dependence of the spin-wave pole amplitude, but does not fully account for the high energy tail at $\mathbf{q}$=(1/2,0) in Fig.~\ref{Fig:e_cuts}(c). Finally, we fitted the complete lineshape (peak and tail) with the generic power-law decay function
\begin{equation}
\chi^{\prime\prime}(\mathbf{q},\omega)=A_{\mathbf{q}} \frac{\theta \left(\omega-\omega_{\mathbf{q}}\right)}{\left( \omega^{2}- \omega^{2}_{\mathbf{q}} \right)^{1-\eta/2}},
\label{Eq:Spinon}
\end{equation}
where $\omega_{\mathbf{q}}$ is an onset energy (peak position), $A_{\mathbf{q}}$ is a wavevector-dependent amplitude factor and $\theta$ is the Heaviside step function. This power-law function is a generalization of the continuum scattering lineshape \cite{Muller1981a} of the 1D Heisenberg AF chain. It gives a good parameterization of the high-energy zone-boundary data, with $\eta=0.93 \pm 0.08$ at $\mathbf{q}$=(1/2,0), where the continuum scattering is strong, and a smaller $\eta=0.47 \pm 0.05$ at $\mathbf{q}$=(1/4,1/4), where the continuum tail is much weaker.  Integrating Eq.~\ref{Eq:Spinon} (for $E<450$~meV) at the two positions we obtain $I(1/2,0)/I(1/4,1/4)=0.90 \pm 0.11$ showing that most of the loss in intensity from the SW pole is in the tail at (1/2,0).  A reduced SW intensity and high energy tail at (1/2,0) have also been observed in the small $t/U$ weak-exchange square-lattice antiferromagnets CFTD \cite{Christensen2007a} and Cu(pz)$_2$(ClO$_4$)$_2$ \cite{Tsyrulin2009a}. Although the tail appears to be weaker in CFTD than in La$_2$CuO$_4$ or Cu(pz)$_2$(ClO$_4$)$_2$.

In general terms, our results show that at the $\mathbf{q}$=(1/2,0) position the spin waves are more strongly coupled to other excitations than at $\mathbf{q}$=(1/4,1/4).  This coupling provides a decay process and therefore damps the spin wave, reducing the peak height and producing the tail. The question is: ``What are these other excitations?''  An interesting possibility is that the continuum is a manifestation of high-energy spinon quasiparticles proposed in theoretical models of the cuprates \cite{Anderson1987a,Hsu1990a,Ho2001a,Sandvik2001a,Lannert2003a,Anderson2004a,Lee2006a}. These assume that N\'eel order co-exists with additional spin correlations with the magnetic state supporting both low-energy spin-wave fluctuations of the N\'eel order parameter as well as distinct high energy spin 1/2 spinon excitations created above a finite energy gap \cite{Ho2001a,Lannert2003a}. Spinons are $S=1/2$ quasiparticles which can move in a strongly fluctuating background.
The anomaly we observe at (1/2,0) may be explained naturally in a model where spinons exist at high energies and have a $d$-wave dispersion \cite{Ho2001a,Lannert2003a} with minima in energy at $\mathbf{q}$=$(\pm 1/4,1/4)$ and $(1/4,\pm 1/4)$.  Under these circumstances, the lower boundary of the two-spinon continuum is lowest in energy at (1/2,0) and significantly higher at (1/4,1/4). This provides a mechanism for the spin waves at (1/2,0) to decay into spinons [with $(1/4,\pm 1/4)$] and those at (1/4,1/4) to be stable.

The qualitatively new features in the collective magnetic excitations observed in the present study are (i) a momentum dependent continuum and (ii) the $\mathbf{q}$-dependence to the intensity of the spin-wave pole. We estimate the total observed moment squared (including the Bragg peak) is $\langle M^2 \rangle$=$1.9 \pm 0.3$~$\mu_B^2$. The continuum scattering accounts for about 29\% of the observed inelastic response. The total moment sum rule \cite{Lorenzana2005a} for $S=1/2$ implies $\langle M^2 \rangle=g^2 \mu_B^2 S(S+1) = 3$~$\mu_B^2$.  We consider two reasons why we fail to observe the full fluctuating moment of the Cu$^{2+}$ ion. Firstly, our experiment is limited in energy range to about 450~meV, thus there may be significant spectral weight outside the energy window of the present experiment. Raman scattering \cite{Sugai1990a} and optical absorption \cite{Perkins1993a} spectra show excitations up to about 750~meV. Recent resonant inelastic X-Ray scattering measurement also show high energy features \cite{Ellis2010a} which appear to be magnetic in origin. The second reason why we may fail to see the full fluctuating moment may be covalency effects \cite{Hubbard1965a,Walters2009a}. The Cu $d_{x{\scriptstyle ^2}-y{\scriptstyle ^2}}$ and O $p_x$ orbitals hybridize to yield the Wannier orbital relevant to superexchange. This will lead to a reduction in the measured response. However, the (at most) 36\% reduction observed in La$_2$CuO$_4$ is substantially less than then the 60\% reduction recently reported in the cuprate chain compound Sr$_{2}$CuO$_{3}$ \cite{Walters2009a}.

Our results have general implications for the cuprates.  Firstly, they show that the collective magnetic excitations of the cuprate parent compounds cannot be fully described in terms of the simple SW excitations of a N\'eel ordered state. Secondly, they demonstrate the existence of considerable high-energy spectral weight above the SW pole. Thus, the ground state of La$_{2}$CuO$_{4}$ contains additional correlations not captured by the N\'eel/SWT picture which may take the form proposed by Anderson and others \cite{Anderson1987a,Hsu1990a,Ho2001a,Sandvik2001a}. Deviations from SWT are strongest at high energies and are signalled by the observation of a $\mathbf{q}$-dependent continuum of magnetic excitations and a strong $\mathbf{q}$-dependent damping of the spin-waves.  The SW pole is most heavily damped at the (1/2,0) [or ($\pi$,0)] position.  This may be understood in a model where the spin waves are coupled to (and decay into) another type of dispersive excitation such as spinons. Angle resolved photoemission spectroscopy (ARPES) studies \cite{Damascelli2003a} show that doping La$_{2}$CuO$_{4}$ or other parent HTS compounds produces an underdoped superconducting state with a $\mathbf{q}$-dependent $d$-wave pseudogap.  The spinon dispersion that can explain the observed spin-wave damping in our experiments has the same form as the pseudogap dispersion, with a minimum at (1/4,1/4).


\end{document}